%Paper: gr-qc/9402037
%From: Gerard 't Hooft <thooft@fys.ruu.nl>
%Date: Mon, 21 Feb 1994 14:36:48 +0100

% Paper written in TEX.
% The 4 figure files, written in Postscript, are added in the end. Strip them
% off (make sure the first characters are %! ), call them th1.ps th2.ps th3.ps
% and th4.ps. With dvips the figures should emerge in the text.

\input epsf

\magnification= \magstep1
\tolerance=1600
\parskip=0pt
\baselineskip= 6 true mm

\font\smallrm=cmr8

\def\a{\alpha}
\def\b{\beta}
 
\def\d{\delta} \def\D{\Delta}

\def\k{\kappa}
\def\l{\lambda} 
\def\m{\mu}
\def\f{\phi} 
\def\n{\nu}
\def\j{\psi} \def\J{\Psi}
\def\r{\rho}
\def\s{\sigma} 

\def\th{\theta}

 \def\W{\Omega}
\def\v{\varphi}

\def\dr{{\rm d}}
\def\cl{\centerline}
\def\pa{\partial}
\def\fn{\footnote{$ ^\dagger $}}
\def\dt{{{\rm d}\over {\rm d}t}}
\def\half{{\textstyle{1\over2}}}
\def\xt{\tilde x}
\def\in{^{\rm in}}
\def\out{^{\rm out}}
\def\pt{\tilde p}
\def\qu{\ {\buildrel ?\over =}\ }

{\nopagenumbers

\vglue 1truecm
\rightline{THU-94/02}

\vfil
\cl{ \bf HORIZON OPERATOR APPROACH TO BLACK HOLE }
\vskip .5 truecm
\cl{{\bf QUANTIZATION.}\fn{\smallrm Presented at "The Black Hole 25
Years After", Santiago, Chile, 17-21 January 1994}}

\vfil
\cl{G. 't Hooft}
\vskip 1 truecm
\cl{Institute for Theoretical Physics}
\cl{University of Utrecht, P.O.Box 80 006}
\cl{3508 TA Utrecht, the Netherlands}
\vfil
{\bf Abstract}
\bigskip
The $S$-matrix Ansatz for the construction of a quantum theory of black holes
is further exploited. We first note that treating the metric tensor $g_{\m\n}$
as an operator rather than a background allows us to use a setting where
information is not lost. But then we also observe that the 'trans-Planckian'
particles (particles with kinetic energies beyond the Planck energy) need to
be addressed. It is now postulated that they can be transformed into
'cis-Planckian particles' (having energies less that the Planck energy). This
requires the existence of a delicate algebra of operators defined at a black
hole horizon. Operators describing ingoing particles are mapped onto operators
describing outgoing ones, preserving their commutator algebra. At short
distance, the transverse gravitational back reaction dictates a discrete
lattice of data points, and at large distance the algebra must reproduce known
interactions of the Standard Model of elementary particles. It is suggested
that further elaboration of these ideas requiring complete agreement with
general relativity and unitarity should lead to severe restrictions concerning
the inter-particle interactions.\vfil\eject}
\pageno=1
\cl{1. INTRODUCTION}
\medskip
The fact that black holes should emit elementary particles at a well-defined
temperature, as can be derived by more-or-less standard techniques in quantum
field theory[1] puts them in the same category as the heavy unstable particle
like solutions that also exist in many other field theories, and therefore one
expects that they exhibit some well-defined spectrum of (excited) states, and
one would expect furthermore that these states should be calculable from the
local laws of physics. Indeed, from the known mass dependence of the
temperature one can readily derive that the density of levels should be given
by[2]
$$\r(M)\,=\,Ce^{4\pi M^2}\ ,\eqno(1.1)$$
in natural units ($G=\hbar=c=1$). Here $C$ is a multiplicative constant.
However, in contrast to the situation in conventional field theories, which are
sufficiently well understood conceptually, this constant $C$ seems to be
fundamentally uncomputable - indeed, a standard calculation following a
background field technique, always gives $C=\infty$, implying a
strictly continuous spectrum.

If this naive result would be correct and the black hole spectrum would indeed
be continuous this would require a radical departure from standard quantum
theory. It would imply that even the tiniest black hole could absorb infinite
amounts of information, and it would never be possible to represent such an
object by a finite component wave function, in contrast to all other known
physical objects. It has been argued that a quantum mechanical wave function
can still be formulated, but only if all possible states in all other
"universes" connected to the black holes by analytic extension of its metric,
would be taken into account. If we were forced to limit ourselves to our own
universe there simply would not exist a Schr\"odinger equation for black
holes. Physicists are now divided mainly in three camps as to what our
attitude towards this problem should probably have to be.

The first proposal, particularly defended by Hawking[3], is that conventional
quantum mechanical behavior may be limited to distance scales much larger than
the Planck scale, but there are fundamental deviations from that at Planck
scale distances. We will have a density matrix $\r(t)$, but in stead of its
usual propagation law
$$\dt \r(t)=-i[H,\r(t)]\ ,\eqno(1.2)$$
he proposes a more general linear evolution law:
$$\dt \r(t)=-i\$\r(t)\ ,\eqno(1.3)$$
where $\$ $ is an arbitrary linear operator acting on all components of the
matrix $\r$. This proposal has been criticised by Banks et al[4], who argued
that conventional conservation laws such as energy conservation will be
violated. We could add to theirs the following consideration. The density
matrix may be seen as describing two universes - which under normal
circumstances do not interact with each other - since it spans a Hilbert space
that is the product of ket- and bra states:
$${\cal H}=\{\r_{ij}\}=\{|\j_i\rangle\}\otimes\{\langle\j_j|\}\ ,\eqno(1.4)$$
which in conventional quantum mechanics evolves according to
$$\dt\r_{ij}(t)=-iH_{ik}\r_{kj}+iH_{jk}\r_{ik}\equiv -i(H^{(1)}-H^{(2)})\r_{ij}
\ .\eqno(1.5)$$
We see that in these two universes energy is defined with opposite signs.
Imagine now that, if a rule such as eq. (1.3) would hold, some sort of
interaction takes place between the two universes:
$$\dt \r(t)=-i(H^{(1)}-H^{(2)}+H^{int})\r(t)\ ,\eqno(1.6)$$
then this would imply that the vacuum state, defined as the zero eigenstate of
$H^{(1)}$ and $H^{(2)}$, would be able to make transitions to any state of the
form $|E\rangle\langle E|$, since this conserves total energy for both spaces
combined. But since phase space for the final states is infinitely larger than
that of the vacuum state the transition would never go backwards to the
vacuum. Or, in terms of kets alone, transitions would be made into higher
energy mixed states, and this would be experienced as an instability of the
vacuum state. Only the absence of an interaction Hamiltonian could protect the
vacuum against such an instability, but this would precisely correspond to a
pure Schr\"odinger equation rather than a $\$ $ matrix evolution law for the
density matrix.

The second proposal often defended is that black holes do absorb all
information thrown into them, such that their spectrum becomes infinitely
degenerate, which then implies that at some lowest energy there must exist an
infinitely degenerate state, called a 'black hole remnant'. These remnants
then cannot decay any further and hence should be absolutely stable. That this
would be the way information is preserved was concluded from calculations in
a two dimensional model by Callan et al[5]. The problem with this proposal
is that these remnants would not form a Bose-Einstein or Fermi-Dirac gas
but rather a 'Boltzman gas', since they are infinitely degenerate. Whatever
process would create tiny black holes at the Planck length would destabilize
the universe thermally, since phase space of a remnant gas is infinitely
larger than that of any other state at the same energy.

The third proposal, preferred by the present author[6, 7], is that the
information is projected into the Hawking radiation. Mathematically,
this assertion means that if two possible initial states for the black
hole were mutually orthogonal, the final states will be orthogonal
also, provided that all Hawking particles are included in the
considerations. Clearly it will be impossible to check such a statement
for macroscopic black holes, so that the latter may in practice be seen
to behave exactly as Hawking found. But at a microscopic level it
implies that there ought to exist an $S$-matrix with poles
corresponding to each metastable black hole configuration. This would
imply that the constant $C$ in eq. (1.1) should be strictly finite.
Note though that the possibility for it being very large or very small, e.g.
$10^{\pm 40}$, is still kept wide open. We have to keep in mind that large
numbers may naturally emerge from quantum gravity!

The problem with this third proposal is that it is in conflict with all
calculations in the linearized field approximation. These are calculations
where quantum operator fields are superimposed onto each other. There are
several equivalent ways to see how this happens[6]. First one may consider
quantum field theory on the background of the metric in the usual
Schwarzschild coordinates $r$ and $t$, where
$$\dr s^2=-\bigg(1-{2M\over r}\bigg)\dr t^2+\bigg(1-{2M\over
r}\bigg)^{-1} \dr r^2+r^2\dr \W^2\ ,\eqno(1.7)$$
and
$\W=(\th,\v)$. The Lagrangian for a scalar field $\f(x)$ with mass $m$ is
$${\cal L}=\half{\sqrt{-g}}\bigg[-\bigg(1-{2M\over r}\bigg)\pa_r\f^2+
\bigg(1-{2M\over r}\bigg)^{-1}\pa_t\f^2-r^{-2}l(l+1)\f^2-
m^2\f^2\bigg]\ .\eqno(1.8)$$
Close to the horizon it becomes manageable in the coordinates
$$\s={\rm log}(r-2M)\quad {\rm and}\quad t\ ,\eqno(1.9)$$
since then
$${\cal L}\dr r\dr t=\half\dr \s\dr t\Big(r^3\pa_t\f^2-r\pa_\s
\f^2-(r-2M)\big[m^2 r^2+l(l+1)\big]\f^2\Big)\ ,\eqno(1.10)$$
At $r\approx 2M$ the Euler-Lagrange equations are readily seen to
produce plane wave solutions, and since $\s$ is bounded at the left
only by $-\infty$ we immediately deduce that the spectrum should be
continuous. Actually the absence of anything like a boundary condition
for linearized fields close to the horizon can also easily be deduced
from arguments in Kruskal or Penrose coordinate frames.

Thus, if we suspect that the information bounces back into the Hawking
radiation we must assume this to be due to interactions. Clearly, as
$e^\s$ plummets below the Planck length the gravitational interactions
become very strong and here the linearized approximation is invalid. We
will therefore always assume that the recovery of information is due to
strong gravitational (back) reactions very near the horizon. A
`boundary condition' very near the horizon has indeed been proposed by
several authors[6, 8].  We do generally assume that Hawking emission
occurs as calculated by Hawking when the black hole is large. In that
case however the hole is apparently only allowed to be a probabilistic
mixture of many states. One must suspect then that general coordinate
invariance only holds for average states, not so much for individual
states.

It is instructive to compare apparent pure-to-mixed transitions with
transitions in a theory with uncertain Hamiltonian. Imagine a
conventional quantum mechanical system with a Hamiltonian depending on
some parameter $\a$:

$$\dt|\J(t)\rangle=-iH(\a)|\J(t)\rangle\ ,\eqno(1.11)$$
where the value for the parameter $\a$ is known to have a probability
distribution $P(\a)\dr \a$. Let the state at $t=0$ be given as a
pure state $\J(0)$. The expectation value of some operator $\cal O$ at
time $t$ is then
$$\langle{\cal O}\rangle=\int\dr \a P(\a)
\langle\J(0)|e^{iH(\a)t}{\cal O} e^{-iH(\a)t}|\J(0)\rangle = {\rm
Tr}\big(\r(t){\cal O}\big)\ ,\eqno(1.12)$$
where $\r(t)$ is easily seen to be the density matrix for a mixed
state.

Apparently the linearized field approximation is tantamount to
admitting some uncertainty in the Hamiltonian. The fact that such an
interpretation did not follow from calculations in some special models
as produced by Ref.[5] may well mean that such special models are not
suitable for an accurate representation of quantum gravity. Quite
generally there seem to be reasons to complain that these models are
not sufficiently explicit when it comes to describing non-perturbative
phenomena such as black holes, a complaint one could also utter against
(super)string theory and even the existing models of quantum gravity in
2+1 dimensions[9] In any case we will observe that a description of
black holes that is completely in accordance to quantum mechanics will
require entirely new physics. Admittedly, adhering to this last option
(i.e. in an accurate theory black holes remain pure) does not make life
easy. Yet we will try to convince the reader that progress from this
starting point looks quite promising.

\bigbreak
\cl{2. THE $S$-MATRIX ANSATZ}
\medskip
\noindent From now on we will make the following assumption ($S$-matrix
Ansatz)[10]:

\smallskip
{\narrower\noindent \it All physical interaction processes, therefore
also all those that involve the creation and subsequent evaporation of
a black hole, can be described by one scattering matrix $S$ relating
the asymptotic outgoing states $|\rm out\rangle$ to the ingoing states
$|\rm in\rangle$.\smallskip}
\noindent We will often use this Ansatz the
following way: we {\it assume} some value for one particular transition
amplitude $\langle\rm out_0|in_0\rangle$, after which we make some long
series of small changes both in the in- and in the out-state. The
effects of the changes are often directly computable. Hence knowing one
matrix element gives access to calculating others, by making use of
known laws of physics. It will turn out that using well-established
physical laws at large distance scales never leads to direct conflicts;
rather, large segments of the $S$-matrix can be computed along these
lines
\midinsert
\epsffile{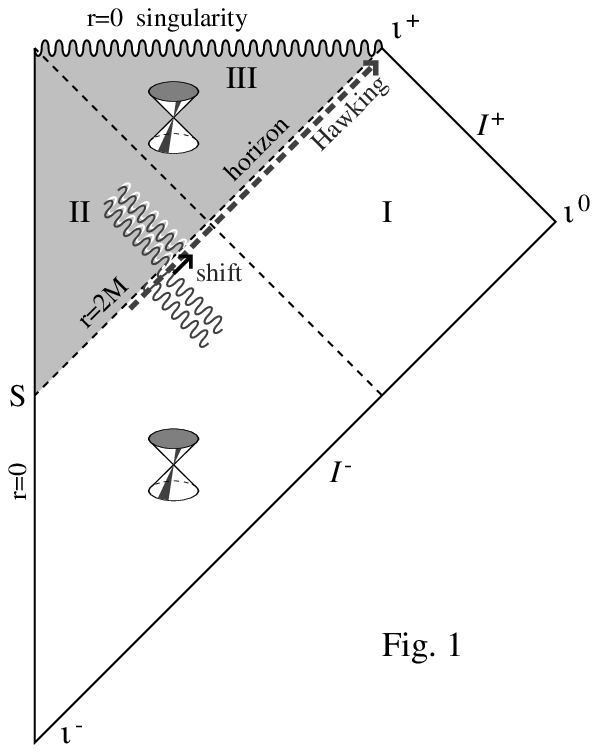}
\cl{\smallrm Penrose diagram for
black hole after formation.}
\endinsert

One often hears the objection that the $S$-matrix Ansatz appears to
violate causality, since matter once fallen through the horizon wll be
spacelike separated from the region where Hawking radiation is seen, so
that this Hawking radiation cannot carry the information. If it did,
this would be an example of "quantum mechanical duplication" -- states
including information about quantum phase factors seem to occur at two
places that are spacelike separated.  This objection however is based
upon the assumption that one could, at least in principle,
independently observe matter that passed through the horizon and
Hawking radiation travelling very close to the horizon. We will refer
to an observer able to do this as a `super obeserver'. We now claim
that such super observers do not exist. To be precise:
{\smallskip\narrower\noindent If we restrict ourselves to Hilbert space
spanned either by all possible asymptotic in-states or by all possible
asymptotic out-states, then operators describing fields within the
horizon do not commute with operators describing Hawking particles --
in fact, all relevant commutators tend to infinity.
\smallskip}
\noindent Clearly this would be sufficient to exclude their independent
observation by some super observer.  But how should the above statement
be compatible with standard wisdom concerning the commutation of
spacelike separated observables? The answer to that is in the
smallprint: we limit ourselves to the Hilbert space spanned by the
asymptotic states. Consider the Penrose diagram of a black hole just
formed, see Fig. 1. In this coordinate frame it is hard to talk about
Hawking radiation at all because it is the standard frame for
describing the single vacuum state at the onset of the horizon. The
Hawking particles live in region $I$, very close to infinity in Fig.1,
and very strongly Lorentz transformed. This we have to keep in mind.
Now consider just any observable operator ${\cal O}_H$ acting on the
visible Hawking particles. The crucial point is that, as seen by
observers in the frame of Fig. 1, these operators create
`trans-Planckian' particles.

By themselves, the trans-Planckian particles cannot be excluded from
the physical Hilbert space of ingoing particles. But then consider
particles passing from region $I$ into region $II$ or $III$, described
by operators ${\cal O}_I$, ${\cal O}_{II}$ and ${\cal O}_{III}$. The
trans-Planckian Hawking particles created by ${\cal O}_H$ affect them
by a {\it gravitational drag} as indicated in the Figure. This shift
depends on the position in the transverse coordinates $(\th,\v)$. It
causes a mismatch between the operators ${\cal O}_I$ on the one hand
and ${\cal O}_{II}$ and ${\cal O}_{III}$ on the other hand. This
mismatch is normally forgotten when ingoing particles in regions $II$
and $III$ are described. Since furthermore the shift depends on the
transverse coordinates its effect cannot be removed by coordinate
redefinitions. Thus we have $${\cal O}_{II}\,{\cal O}_H={\cal
O}_H\,{\cal O}_{II,\ \rm shifted}\ .\eqno(2.1)$$

At first sight the need to perform the shift in eq. (2.1) may perhaps
not be obvious. The point is however that the Hilbert space we chose to
work in is ${\cal H}_I$, the one spanned by the states in $I$, not
${\cal H}_{II}$ which is spanned by the states in $II$ and $III$. If
the latter were the case we could have used the local coordinate frame
of $II$ and $III$ and there would be complete commutation. However, the
Hilbert space space ${\cal H}_{II}$ does not allow the action of ${\cal
O}_H$ without generating a {\it white hole} at the onset of the
horizon, $S$ in Fig. 1. {\it The absence of a white hole is one of the
most essential elements of the $S$-matrix Ansatz.} We observe that the
non-commutation as described by eq. (2.1) is linked to the need to
describe the space-time metric as a quantum operator rather than a
c-number background. This was also emphasized in Ref. [11]. It is now not
difficult to see that the commutator diverges as $t\rightarrow\infty$,
where $t$ is time parameter in ${\cal O}_H$, being the moment Hawking
radiation is studied by the external observer.
		\vfil   % \vfil perhaps to be removed later
\bigbreak
\cl{3. THE NON-SINGULAR METRIC}
\nobreak\medskip\nobreak
How should one take the quantum nature of the metric into account? One
proposal was made in Ref. [12]. If the in-state is well-specified the
metric can be treated as a c-number\fn{\smallrm We treat the metric
operator $\hat g_{\m\n}({\bf x},t)$ as if the in-state is an
eigenvector for it at all $({\bf x},t)$ during the collapse, but this
of course is not true. $g_{\m\n}$ and $\dot g_{\m\n}$ obey canonical
commutation relations. However, these can be regarded as the fields
describing gravitons, which will be considered as small fluctuations.
In the present argument it is the large scale fluctuations of
$g_{\m\n}$ that we are describing and trying to keep under control.}
until the horizon is reached. Now if we work with the $S$-matrix Ansatz
we can also specify the out-state such that also the metric during
evaporation is well-determined as soon as the white hole's past-horizon
is left behind. The philosophy of the $S$-matrix Ansatz is to consider
only small changes both in the in-states and the out-states. To do this
we only need the metric $g_{\m\n}({\bf x},t)$ as defined by

$$g_{\m\n}={\langle {\rm out}|{\hat g}_{\m\n}|{\rm in}\rangle\over
\langle{\rm out}|\rm in\rangle}\ ,\eqno(3.1)$$

\noindent which is well-defined both during the initial and the final
phases. It is therefore proposed to glue the metrics for the in-state
and the out-state together, as is depicted in Fig. 2. The relevance of
in/out matrix elements of observables as in eq. (3.1) was discussed by
Aharonov and Vaidman [13]. They call $g_{\m\n}$ as defined in
this equation the `weak value' of the operator ${\hat g}_{\m\n}$.

\medskip
\midinsert
\epsffile{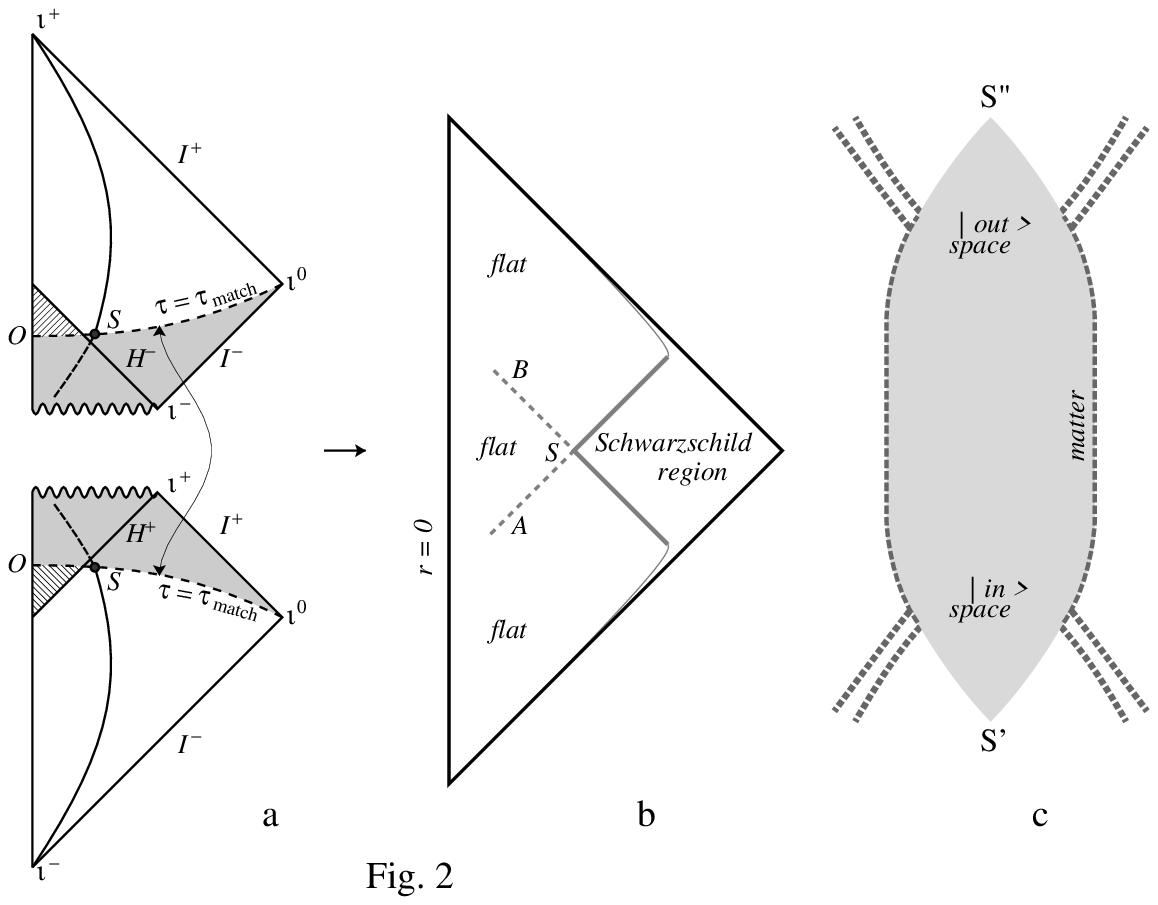} \noindent\narrower{\smallrm a) The metric
generated by $g_{\m\n}({\bf x},t)$ as it acts on the {\rm in} state
(below) and the {\rm out} state (above). b) The glued metric as matrix
element between {\rm out} and {\rm in} state. The shaded regions in (a)
are removed since they would become ambiguous. The regions with the
slanted lines in (a) are also glued together, so that in this metric
information is restored. c) The resulting metric in less deformed
coordinates.} \endinsert

The simplest situation to be discussed first is the case where the
in-state  is chosen to contain a single radially symmetric shell of
matter moving inwards, and the out state to contain also a single shell
of matter, now moving outwards. Although this out-state does not
represent Hawking radiation we expect this amplitude to be of the same
order of magnitude as the amplitudes for Hawking radiation, the only
reason for this particular out-state to be less likely being its much
smaller phase space factor. Within the shells space-time is flat, and
this allows us to glue the metrics of in-space and out-space together
also at the inside.

The price one pays for obtaining this singularity-free overall metric
(Fig. 2b) is that at the point where the in- and out-shells meet,
labled $S$ in Fig. 2a and b, the curvature is very large. A physical
description of this point is to say that a very violent explosion takes
place there, sending ingoing matter back out; the curvature results
from the huge stress tensor that is needed for this. in the limit of
infinitely thin shells we do have a singularity at $S$ but it is a mild
one: a {\it cusp} singularity. It is surrounded by nearly-flat
space-time such that a space-time journey looping around $S$ produces a
Lorentz transformation.

One can now proceed to do quantum field theory in this topologically
trival space-time. Certainly no "information gets lost". The $S$-matrix
Ansatz does require that we consider weak and low-energy fields only,
and this will be an important restriction as we will see. At first
sight such a restriction does not seem to be worrisome. We now should
be able to consider small perturbations $\d \langle{\rm out}|$ and
$\d|{\rm in}\rangle$ and compute the ampltudes for the perturbed
states\fn{\smallrm Note that in terms of distances in Hilbert space the
perturbations need not be small; the perturbed states are generally
orthogonal to the original ones; the perturbations are small in the
sense that they have little gravitational effects.}. Field theory shows
interesting effects near $S$. The cusp is a particle producer: if we
have a vacuum preceeding $S$ the state behind $S$ will contain
particles (in a quantum mechanically coherent state). Details of the
mechanism are not difficult to compute; we refer to Ref. [12]. The result
is that the intensity of the particle production by a cusp in the
$x\,y$ direction is given by
$${\dr N\over \dr p_z \dr^2\xt}={1\over p_0}\Big({\a\over{\rm
th}{\a\over2}}-2\Big)\ ,\eqno(3.2)$$
where $\a$ is the generator of the Lorentz boost surrounding the cusp.
We see that this expression vanishes as $\a\rightarrow\infty$ and it
grows linearly in $|\a|$ as $|\a|\rightarrow\infty$.

\midinsert
\epsffile{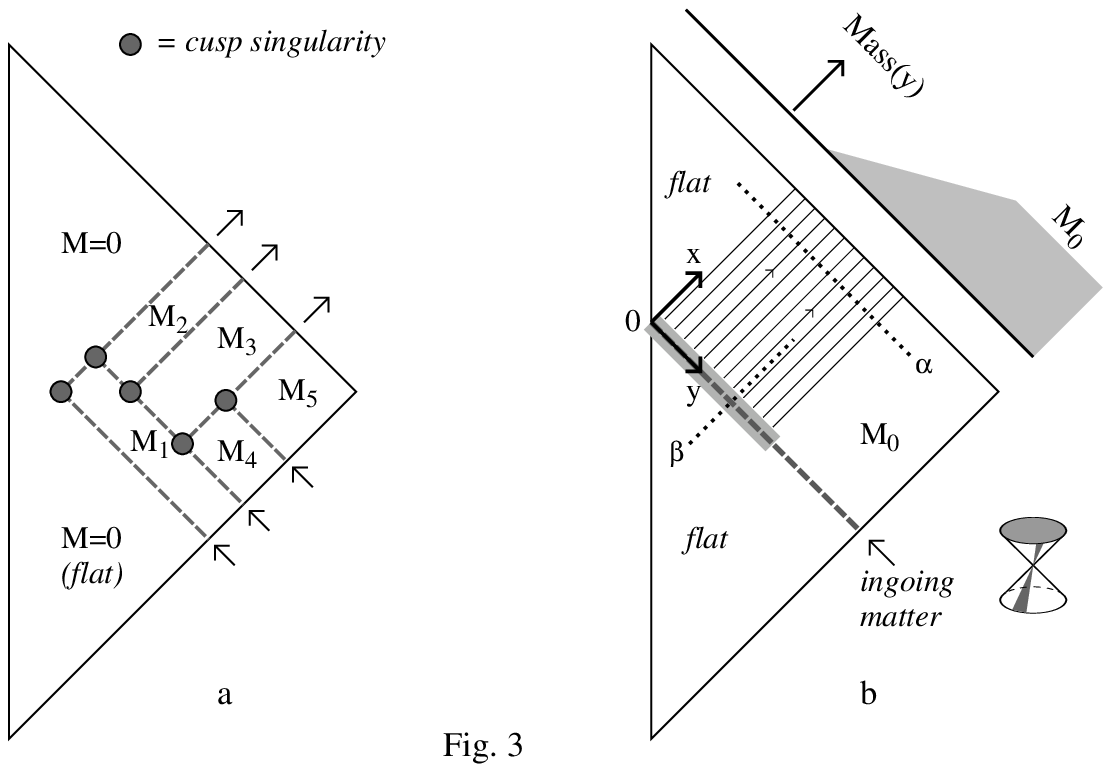}
\narrower{\smallrm a) Process with 3 shells in the in-state and 3
shells in the out state. Here we have $0<M_1<M_2<M_3<M_4<M_5$ . b) One
shell in and a near continuum of shells out. On top the residual mass
as a function of the $y$ coordinate. Shaded: region with strong
curvature. Dotted lines $\a$ and $\b$: see text.}
\endinsert

As a next step one might consider having several shells of matter both
in the in-state and in the out-state. The result is sketched in Fig.
3.  In the regions $i$ in between the black hole has a mass $M_i$. The
most natural prescription seems to be to extend the mass shells as far
to the origin as possible without creating negative mass values
anywhere. Again the $S$-matrix Ansatz implies unconventional physics at
the intersection points, but this was in the postulate that the
corresponding amplitude was unequal to zero. Again we can then consider
doing quantum field theory in this space-time.

Finally one might hope to consider the `most interesting' case, which
is the choice of physically realistic Hawking radiation in the
out-state (Fig. 3b). The in-state can be kept at its simplest, a single
shell. Doing quantum field theory here might help us to unearth more
precise values for all sorts of amplitudes. Now the $y$ coordinate near
the upper right corner of Fig. 3b can be mapped onto the time
coordinate $t$ for the distant observer. It is easy to see that during
the radiation in a good approximation the mass will vary linearly as a
function of $y$. Since time is proportional to $M^3$ we find a cubic
relation between $y$ and $t$.

Again, our philosophy is to use this metric as a background for a
quantum field theory in order to compute "neighboring" amplitudes. In
practice however there turn out to be important difficulties.  Figures
2a,b and 3a,b are Penrose diagrams. This means that the local speed of
light is easy to read off; light propagates at a maximum of $45^\circ$.
However, distances cannot be derived directly from the figures; they
are very distorted. Write the metric as
$$\dr s^2=A\dr x \dr y +r^2\dr\W^2\ ,\eqno(3.3)$$
where $A$ and $r$ depend on the coordinates. The function $r$ is quite
regular, having a natural zero at the origin. But $A$ behaves very
wildly, and this turns out to be a real problem.  Demanding the line
$r=0$ to be vertical puts a constraint on the lightcone coordinates $x$
and $y$ which removes some of the arbitrariness in $A$ that is due to
redefinitions of $x$ and/or $y$. In the Appendix the metric of Fig. 3b
is calculated. The result is given in eqs. (A.14) and (A.15). From
these one can read off the red- and blueshift factors associated to
trajectories through this space-time. The blueshift along the
trajectory labled $\a$ in Fig. 3b is found to be:
$${\pa r\over \pa x}\rightarrow e^{{M_0}^2/4\l}\ .\eqno(3.4)$$
Following the trajectory $\b$ one finds a redshift of only
$${\pa y\over \pa r}\rightarrow {\l\over M_0M^2}\ .\eqno(3.5)$$

Both of these results present us with a problem. The redshift (3.5)
seems to imply that the information from ingoing material will be
spread over extremely long wavelengths in the outgoing radiation. But
(3.4) is even more disastrous. Information entering via trajectory $\a$
will be turned into extreme trans-Planckian particles. Indeed the
blueshift is so extreme that the energies of outgoing particles would
tend to surpass quickly the energy of the entire universe, which of
course would be nonsence. The $S$-matrix Ansatz forces us to omit such
final states; yet we seem to obtain that precisely here the information
of ingoing material goes.

A -- possibly related -- problem is the observation that when we apply
field theory in these non-singular space-times we will of course
respect all symmetries of the initial quantum field model, including
possible baryon number conservation. Now surely the background metric
does violate baryon number conservation (in most interesting cases),
but now we see that the changes we can consider in the in- and
out-states will not bring us from one channel with baryon violation
$B_{\rm out}-B_{\rm in}=\D B$ to any other channel with different $\D
B$. This wil make it difficult for us to study the details of any
baryon number violating phenomena in black holes, even though our
theory will permit these violations.
\bigbreak
\cl{4. TRANS-PLANCKIAN TO CIS-PLANCKIAN MAPPING: OPERATOR ALGEBRA}
\medskip
{}From the previous section we conclude that a unitary field theory in a
non-singular space-time may lead to a unitary $S$-matrix only if we
allow for states that should actually be considered inadmissible:
particles with energies far beyond the Planck mass, even far beyond the
black hole mass. Our Ansatz forces us to limit ourselves to particles
with energies up to the Planck mass but not much beyond. It is here
where we are really forced to modify the laws of physics as we know
them. At the points $A$ and $B$ in Fig. 2b one would normally expect
only near-vacuum states whereas our field theories seem to generate
strongly blueshifted particles there. It seems to be inevitable that a
transformation law has to exist allowing us to transform states with
particles much beyond the Planck energy into states where all particles
have less than a Planck unit of energy. Again the $S$-matrix Ansatz can
be helpful here to obtain this transformation rule.

This procedure has been described in Ref. [10]. Making use of the fact
that ingoing material {\it always} affects outgoing waves via
gravitational interactions (if not other interactions as well) we can
consider a small change $\d_{\rm in}$ in the ingoing state $|\rm in
\rangle$, having a momentum distribution $\d p\in(\th,\v)$. The
geodesics of outgoing particles will be shifted, as one can easily
derive [10], and the shift $\d y$ in the $y$ coordinate is given by
$$\d y(\W)=\int \dr\W' G(\W,\W')\d p\in(\W')\ ,\eqno(4.1)$$ where $\W$
stands for $(\th,\v)$ and $G(\W,\W')$ is a Green function on the
transverse coordinates, determined by the equation $$\big(1-\pa_\W
^2\big)G(\W,\W')=C\d^2(\W,\W')\ .\eqno(4.2)$$ Here $C$ is a known
constant depending on the units used and the gravitational constant.
The solution to this equation is $$G(\W,\W')=\k\int_\th^{2\pi-\th}\dr
z\big(\cos\th-\cos z\big)^{-\half}\,e^{-\half\sqrt{3} z}\ ,\eqno(4.3)$$
where $\k$ is related to $C$, and $\th$ is the angle between $\W$ and
$\W'$. It is important however to stress that eq. (4.1) is an
approximation: it was assumed that the ingoing particles were massless
and had negligible transverse momenta. Also all non-gravitational
interactions were ignored.

Since the shift operator is
$$\exp\big(i\int\dr^2\W p\out(\W)\d y(\W)\big)\ ,\eqno(4.4)$$
we notice that all information concerning the in-states being used is
the momentum distribution $p\in(\W)$. Also the outgoing states are
only distinguished by their momentum distribution $p\out(\W)$. We
now make the following essential step: we assume that {\it all}
information concerning these states is in these momentum distributions.
{}From a physical point of view this seems to be reasonable. If we were
able to determine $p(\W)$ with a Planckian resolution we really would
have a lot of information which seems to be more than sufficient in
practical situations. In physics at long distance scales our assumption
therefore seems to imply little restrictions; at small distances
however this restriction is crucial.

Let us for simplicity now turn to Rindler space in stead of the
original black hole. This is not much else than replacing the angular
variables $\W=(\th,\v)$ by $\xt=(x^1,x^2)$.  Let us describe the
in-state as $|p\in(\xt)\rangle$ and introduce its functional
Fourier transform $|u\in(\xt)\rangle$ by demanding
$$\langle u\in|p\in\rangle = \exp\big[i\int\dr^2\tilde
x\,p\in(\xt)\,u\in(\xt)\big]\ .\eqno(4.5)$$
As was shown in ref.[10] the $S$-matrix is generated by giving the
following inner product between the out-states and the in-states:
$$\langle p\out|p\in\rangle = {\cal
N}\exp\big[-i\int\dr^2\xt\dr^2\xt'\,p\out(\tilde
x)f(\xt-\xt')p\in(\xt')\big]\ ,\eqno(4.6)$$
where $\cal N$ is a normalization factor and $f$ is now the Green
function defined by
$$\tilde\pa^2f(\xt)=-\d^2(\xt)\ ,\eqno(4.7)$$
in units were $4\pi G=1$ (notice that this normalization is now
different from what was used in the earlier sections; in black hole
descriptions one usually avoids the $4\pi$).

For black holes this definition of the inner products just generates
the $S$-matrix. However, if we look at these identities from the point
of view of local observers at the horizon, they provide for the
required mapping between trans-Planckian and cis-Planckian particles.

It is now very convenient to introduce operators describing in- or
outgoing particles. The operators $p\in(\xt)$ and $p\out(\xt)$ just
measure these quantities. The operators $u\in$ and $u\out$ are the
canonically conjugated operators. They satisfy
$$\eqalign{[p\in(\xt),p\in(\xt')]&=0\ ;\cr
[p\in(\xt),u\in(\xt')]&=-i\d^2(\xt-\xt')\ ;\cr
[p\out(\xt),p\out(\xt')]&=0\ ;\cr
[p\out(\xt),u\out(\xt')]&=-i\d^2(\xt-\xt')\ .\cr}\eqno(4.8)$$
Furthermore we have
$$u\out(\xt)=\int\dr^2\xt'f(\xt-\xt')p\in(\xt')\ ,\eqno(4.9)$$
from which
$$\eqalign{[p\out(\xt),p\in(\xt')]&=i\tilde\pa^2\d^2(\xt-\xt')\ ;\cr
[u\out(\xt),u\in(\xt')]&=-if(\xt-\xt')\ ,\cr}\eqno(4.10)$$
so that
$$\tilde\pa^2 u\out=-p\in\ ,\quad\tilde\pa^2 u\in=p\out\ .\eqno(4.11)$$
Apparently our equations are very symmetric under time reversal.
So-far this algebra has been used in Ref. [10], where it is also
explained how remarkably the resulting functional integral expressions
for the $S$-matrix resemble the ones in string theory.

There are however two serious shortcomings stil present in these
expressions:\smallskip \item{\it i)} The {\it spectrum} of states is
still continuous, sine the $p$ and $u$ are continuous operators, and
also since they depend on the contiuous transverse coordinates
$\xt$.\item{\it ii)} The relations between representations of our
operator algebra on the horizon on the one hand and ordinary Fock space
of elementary particles in the surrounding four-dimensional space-time
on the other, seem to be difficult to recover. Fock space after all is
not only described by the momentum distribution but also by the
essential particle counting operators. It seems that we replaced the
usual Fock space by a space where particles approaching each other
extremely closely in the transverse coordinates become
indistinguishable, even if they stay far apart in the longitudinal
(light cone) coordinates.\smallskip

In order to cure these shortcomings we observe that the above algebra
is not yet accurate.  Not only did we ignore all non-gravitational
interactions, we also have not yet taken into account the transverse
parts of the gravitational interactions. These transverse parts become
important as soon as particles approach each other at distance scales
shorter than the Planck length in the transverse directions. The
inclusion of non-gravitational forces is in principle straightforward.
In particular the electro-magnetic force can be added in an elegant
way[14]. With all these new forces we do also obtain new degrees of
freedom at small distances, such as the charge density $\r(\xt)$ with
its canonically associated variable: a periodic fifth Kaluza-Klein
coordinate[14]. It is the transverse gravitational force that will
provide the answer to the questions raised above. We have to note that
there will also be a momentum distribution in the sideways direction,
$\pt\in(\xt)$ and $\pt\out(\xt)$. These operators generate sideways
displacements, and as such can be easily defined:
$$\pt\in(\xt)=p\in(\xt)\tilde\pa
u\in(\xt)\ ,\quad\pt\out(\xt)=p\out(\xt)\tilde\pa
u\out(\xt)\ .\eqno(4.12)$$
The commutation rules with the other operators, taking either all
operators referring to the in-space, or all referring to out-space, are:
$$\eqalign{[\pt_(\xt),u(\xt')]&=-i\d^2(\xt-\xt')\tilde\pa u(\xt)\ ;\cr
[\pt(\xt),p(\xt')]&=ip(\xt)\tilde\pa \d^2(\xt-\xt')\ ;\cr
[\pt_i(\xt),\pt_j(\xt')]&=-\pt_j(\xt)\pa_i\d^2(\xt-\xt')-
i\pt_i(\xt')\pa_j\d^2(\xt-\xt')\ .\cr}\eqno(4.13)$$

Now if an incoming particle moves in with a sideways component the
gravitational shift it produces also has a sideways component.
Therefore the equations (4.9)--(4.11) should actually be seen as the
third components of vector equations. Hence one could expect to have,
analogously to (4.10), also
$$[\pt\out_i(\xt),\pt\in(\xt')]\qu
i\tilde\pa^2\d^2(\xt-\xt')\d_{ij}\ .\eqno(4.14)$$
However, this cannot be right. Eq. (4.14) in combination with the previous
commutators do not obey Jacobi's identities. So we have to ask how we
can alter these equations such that at long distance scales the effects
ingoing particles have on outgoing ones are still described in
accordance with the $S$-matrix Ansatz while at short distances the
Jacobi identities are restored. This now may be regarded as a
challenging puzzle. We think that the present paper gave a rough
outline for the rules of the game, but a completely satisfactory answer
has not yet been found. We do have a suggestion as to what direction
one could consider going.
\bigbreak

\cl{ 5. DISCRETENESS ON THE HORIZON}
\medskip
Instead of a continuous two dimensional space of transverse points
$\{\xt\}$ suppose that these points are discretized. Thus we replace
the points $\xt$ by points indicated by single capital letters $A,\ B$,
\dots, and assume that al avery point $A$ we have operators
${x^i}\in_A$, ${x^i}\out_A$, ${p_i}\in_A$ and ${p_i}\out_A$, where
$i=1,\,2,\,3$. The value $i=3$ corresponds to the former $z$ components
and $i=1,\,2$ describes the position in transverse space. We can then
introduce ordinary commutators:
$$[{x^i}\in_A,\,{p_j}\in_B]=i\d^i_j\d_{AB}\ ,\quad\it etc.\eqno(5.1)$$
We now wish to find a relation between the in- and out-operators that
would reproduce Eq. (4.11) in the continuum limit. To do this we
introduce a {\it lattice} on our horizon. What this really means is
that links are defined between points, and these links then define
which points are to be considered as being each other's neighbors,
whereas the minimal number of links between any pair of points $A$ and
$B$ defines a transverse distance between $A$ and $B$. The lattice is
two-dimensional. It is reasonable to demand that links do not cross
each other: the lattice is planar. We see no reason yet to restrict
ourselves to special kinds of lattices such as a square or triangular
lattice. More satisfactory will be the `random' lattice, see Fig. 4.
\midinsert
\epsffile{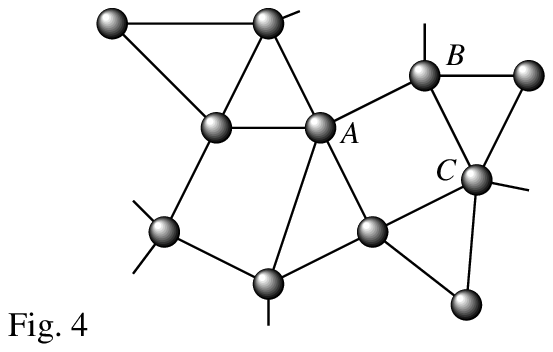}
\cl{\smallrm Lattice at the horizon.}
\endinsert

It is suggestive to replace eq. (4.11), or $p\out =\tilde\pa^2 u\in$, by
$${p_i}\out_A=-{x^i}\in_A+\langle{x^i}\in\rangle_{\rm
linked\,to\,A}\ ,\eqno(5.2)$$
where the average is over the neighbors of $A$ only, with possible
weight factors $C_{AB}$:
$$\eqalign{{p_i}\out_A&=\sum_B C_{AB}{x^i}\in_B\ ,\cr
C_{AA}=-1\ &,\quad \sum_B C_{AB}=0\ .\cr}\eqno(5.3)$$
The coefficients $C_{AB}$ tell us whether points $A$ and $B$ are
neighbors. In that case $$C_{AB}\approx 1/N\ ,\eqno(5.4)$$  $N$ being the
number of neighbors. If $A$ and $B$ are not neighbors $C_{AB}$
vanishes.

Consequently we have
$$[{p_i}\out_A\,,{p_j}\in_B]=i\d_{ij}C_{AB}\ ,\eqno(5.5)$$
which is beautifully time reversally symmetric provided that in
addition to (5.3) we have
$$\sum_A C_{AB}=0\ .\eqno(5.6)$$
This implies that if our lattice is not a regular one (such that the
number of neighbors is variable) the coefficients $C_{AB}$ will have to
be chosen in a more complicated way than just $1/N$, which is why the
`approximatively' sign $\approx$ is used in eq. (5.4).

The radical difference between the algebra of this section with the
previous one is not only the discreteness, but also the fact that there
is now {\it additional} information on the surface: the details of the
lattice structure. Coarse graining adds information, there is little to
be done about that. Allowing the lattice to have a random structure is
mandatory if we want to give the horizon of a finite size black hole
the necessary $S_2$ topology. But this is not a high price to pay; it
implies extra local information of the same type as the information
that would be added if we consider new sorts on interaction at
extremely high energies. This is because the information of the lattice
structure is strictly local.

Finally, we now suspect that our discrete algebra indeed gives the
black hole a discrete spectrum. This is not yet completely evident
since the spectrum of the $p\in$ operators alone, or the $p\out$
operators all by themselves, is still continuous. But if we try to
`localize' the black hole by putting some constraints on the in- and
the out-operators in combination, for instance if we try to localize
the position of the horizon by demanding ${{\bf x}\in}^2+{{\bf
x}\out}^2<R^2\,$, where $R$ is some limiting size, then clearly only a
finite number of states satisfy this constraint since our system in all
respects then behaves as a harmonic oscillator\fn{The resemblance to a
harmonically vibrating membrane becomes even more evident if we make a
restriction in momentum space: give a bound to
$\sum_A({p\in_A}^2+{p\out_A}^2)$.}.

needless to state that many questions are left unanswered. It should be
possible to obtain more information about the rules at a black hole
horizon by exploiting more of the physical information we already have
from the Standard Model concerning the Hilbert space we are trying to
construct. This way we might get a handle on questions concerning the
choice of the coefficients $C_{AB}$ and any other possible local
degrees of freedom.

\bigbreak
\cl{ APPENDIX: METRIC FOR IN- AND OUTGOING SHELLS}\smallskip\cl{OF
LIGHTLIKE MATERIAL.} \medskip
The metric of Fig. 2, produced by discrete shells of matter which are
themselves delta-distributed can of course be written down in closed
form: these space-times consist of various pieces of Schwarzschild
geometry glued together on lightlike seams, such that at least the
transverse part of the metric, given by $r(x,y)$, is continuous there.
Everywhere except at the conical singularities we have that the
components $T_{xy}$, $T_{\th\th}$, $T_{\th\v}$ and $T_{\v\v}$ of the
stress-energy tensor vanish. At the conical singularities the
transverse part of $T_{ij}$ has a delta-distribution.

In this appendix we remove the conical simgularities and consider the
continuum limit of the case of radially symmetric ingoing and outgoing
lightlike shells passing through each other. In the longitudinal
direction we use the lightcone coordinates $x$ and $y$. The metric is
then
$$\dr s^2=A(x,y)\dr x \dr y +r^2(x,y)\dr\W^2\ ,\eqno(A.1)$$
In the discrete case there was a well-defined Schwarzschild mass
parameter everywhere between the shells. In the continuum limit we
still have such a parameter $M(x,y)$. It is defined by
$$A\equiv{2r\, r_x r_y\over r-2M}\ ,\eqno(A.2)$$
where $r_x$ stands for $\pa r/\pa x$. The conditions
$T_{xy}=T_{\th\th}=0$, corresponding to $R_{xy}=R_{\th\th}=0$ imply
$$r_{xy}={2Mr_xr_y\over r(r-2M)}\ ;\quad M_{xy}=-{2M_xM_y\over
r-2M}\ ,\eqno(A.3)$$
and these can be integrated to give
$$2M_xr_x=g(x)\Big(1-{2M\over r}\Big)\ ;\quad
2M_yr_y=h(y)\Big(1-{2M\over r}\Big)\ ,\eqno(A.4)$$
where $g(x)$ and $h(y)$ are arbitrary functions. A physical constraint
is that $g$ and $h$ should be non-negative since
$$T_{xx}={2g(x)\over r^2}\ ,\quad T_{yy}={2h(y)\over r^2}\ .\eqno(A.5)$$

By redefining the coordinates $x$ and $y$ one could normalize the
functions $g$ and $h$ to be equal to one. Physically then the
coordinates $x$ and $y$ measure the amount of material entering and
leaving the hole. The equations (A.4) however cannot be solved in
the completely general case.

In Fig. 3 however we have one delta-distributed ingoing shell and a
fairly arbitrary outgoing shell. Outside the ingoing shell we then have
$$g(x)=0\ .\eqno(A.6)$$
The physically interesting special case is then
$$M_x=0\ ;\quad M=M(y)\ .\eqno(A.7)$$
Let us now assume that far away from the hole the mass loss per unit
of time is given:
$${\dr M\over\dr t}=-F(M)=-{\l\over M^2}\ ,\eqno(A.8)$$
the latter being the expected intensity of the Hawking radiation. $\l$
is of order one and will here be taken to be constant although in
reality it will depend slightly on $M$. Since the radiation goes with
the speed of light we can take at large $r$
$$-{\pa M\over \pa t}\Big|_r\rightarrow {\pa M\over\pa
r}\Big|_t\rightarrow {\pa M/\pa y\over\pa r/\pa y}\ ,\eqno(A.9)$$
leading to
$${\l\over M^2}\rightarrow {2{M_y}^2\over h(y)}\ .\eqno(A.10)$$
Let us normalize the $y$ coordinate by (see Fig. 3b)
$$M(y)=M_0y\ ,\eqno(A.11)$$
so that the range of the $y$ coordinate is [0,1], and $M_0$ is the
initial mass. From eqs (A.10) and (A.4) we get
$$h(y)={2M^2{M_y}^2\over\l}\ ;\quad {\pa r\over\pa
M}={M^2\over\l}\Big(1-{2M\over r}\Big)\ .\eqno(A.12)$$

Close to the horizon, where $r\approx 2M$, the solution to this
equation is $$r(M)=2M+{4\l\over M} +Ce^{M^2/4\l}+\ \it higher\
orders.\eqno(A.13)$$ The $x$ coordinate can now be normalized with the
integration constant $C$.  We do this such that at $y=1$ the $x$
dependence is regular:  $$r(x,y)=2M_0y+{4\l\over
M_0y}+x\,e^{{M_0}^2(y^2-1)/4\l}+\ \it small\ corrections.\eqno(A.14)$$
With eq. (A.1) this leads to
$$A(x,y) \approx {2{M_0}^3y^2\over \l}e^{{M_0}^2(y^2-1)/4\l}\ .
\eqno(A.15)$$
It is the very strong $y$ dependence of these expressions that is
further discussed in section~3.

\bigbreak
\cl{REFERENCES}\medskip
\item{1.}S.W.~Hawking, {\it Commun. Math. Phys.} {\bf 43} (1975) 199;
W.G. Unruh, {\it Phys. Rev.} {\bf D 14} (1976) 870; R.M.~Wald, {\it
Commun. Math. Phys.} {\bf 45} (1975), 9; J.B.~Hartle and S.W.~Hawking,
{\it Phys.  Rev.} {\bf D 13} (1976) 2188.
 \item{2.}J.D.~Bekenstein, {\it Phys. Rev.} {\bf D 5} (1972) 1239;
 G.~'t~Hooft, {\it in} ``Between Science and Technology'', A.~Sarlemijn
 and P.~Kroes (Editors), Elsevier Science Publ. B.V. (North Holland),
 1990, p.77 \item{3.} S.W.~Hawking, {\it Phys. Rev.} {\bf D 14} (1976)
2460; {\it Commun. Math. Phys.} {\bf 87} (1982) 395.
\item{4.}T.~Banks, L.~Susskind and M.E.~Peskin, {\it Nucl. Phys.} {\bf
B 244} (1984) 125; J.~Ellis, J.~Hagelin, D.V.~Nanopoulos and
M.~Srednicki, {\it Nucl.  Phys.} {\bf B 241} (1984) 381.  \item{5.}
C.G.~Callen, S.B.~Giddings, J.A.~Harvey and A.~Strominger, {\it Phys.
Rev.} {\bf D 45} (1992) R1005.  \item{6.}  G.~'t~Hooft, {\it
Nucl.Phys.} {\bf B 256} (1985) 727.  \item{7.}  G.~'t~Hooft, Acta Phys.
Polon. B 19 (1988) 187 \item{8.}  L.~Susskind, L.~Thorlacius and
J.~Uglum, {\it ``The Stretched Horizon and Black Hole
Complementarity''}, Stanford preprint SU-ITP-93-15.
\item{9.}E.~Witten, Nucl.Phys. B 311 (1988) 46; S.~Carlip, Nucl.Phys. B
324 (1989) 106; G.~'t~Hooft, Class. Quantum Grav. 10 (1993) 1653.
\item{10.}G.~'t~Hooft, {\it Nucl. Phys.}{\bf B 335} (1990) 138; {\it
Physica Scripta} {\bf T 36} (1991) 247 \item{11.}K.~Schoutens, E.~and
H.~Verlinde, CERN preprint CERN-TH.7142/94, PUPT-1441, Jan. 1994,
hep-th/9401081 \item{12.}C.R.~Stephens, G.~'t~Hooft and B.F.~Whiting,
{\it Black Hole Evaporation without Information Loss}, Utrecht prepr.
THU-93/20; UF-RAP-93-11; gr-qc/9310006; {\it Class. Quantum Grav.}, to
be publ.; see also G.~'t~Hooft {\it in} Proceedings of the Int. Conf.
on Fundamental  Aspects of Quantum Theory, to celebrate the 60th
birthday of Yakir Aharonov, December 10-12, 1992, Columbia, SC., to be
pub., \item{13.} Y.~Aharonov and L.~Vaidman, {\it J. Phys.} {\bf A 24}
(1991) 2315 \item{14.} G.~'t~Hooft, {\it in} ``Black Hole Physics'',
V.~De~Sabbata and Z.~Zhang (editors), 1992 Kluwer Academic Publ., p.
381
\end